\documentclass[twocolumn]{aastex63}

\received{\today}
\revised{}
\accepted{}

\usepackage{placeins}
\usepackage{amssymb, amsmath}
\usepackage[normalem]{ulem}
\usepackage{tabularx}
\newcommand{\Ms}{M_{\odot}}

\newcommand\T{\rule{0pt}{2.0ex}}       
\newcommand\B{\rule[-1.3ex]{0pt}{0pt}} 

\shorttitle{Characterizing quasi-universality in post-merger gravitational waves}
\shortauthors{C.A. Raithel \& E.R. Most}

\begin{document}

\title{Characterizing the breakdown of quasi-universality in the post-merger gravitational waves from binary neutron star mergers}

\author[0000-0002-1798-6668]{Carolyn A. Raithel}
\email{craithel@ias.edu}
\author[0000-0002-0491-1210]{Elias R. Most}
\email{emost@princeton.edu}
\affiliation{School of Natural Sciences, Institute for Advanced Study, 1 Einstein Drive, Princeton, NJ 08540, USA}
\affiliation{Princeton Center for Theoretical Science, Jadwin Hall, Princeton University, Princeton, NJ 08544, USA}
\affiliation{Princeton Gravity Initiative, Jadwin Hall, Princeton University, Princeton, NJ 08544, USA}
\email{Both authors have contributed equally to this work.}

\begin{abstract}
The post-merger gravitational wave (GW) emission from a binary neutron star merger is expected to provide exciting new
constraints on the dense-matter equation of state (EoS). 
Such constraints rely, by and large, on the existence of quasi-universal relations,
which relate the peak frequencies of the post-merger GW spectrum to properties of the neutron
star structure in a model-independent way. 
In this work, we report on violations of existing quasi-universal relations between the peak spectral frequency, $f_2$, and the stellar radius, for EoSs models with backwards-bending slopes in their mass-radius relations (such that the radius increases at high masses). 
The violations are extreme, with variations in $f_2$ of up to $\sim600$~Hz between EoSs that predict 
the same radius for a 1.4~$\Ms$ neutron star, but that have significantly different radii at higher masses.
Quasi-universality can be restored 
by adding in a second parameter to the fitting formulae that depends on the slope of the mass-radius curve. We further find strong evidence that quasi-universality is
 better maintained for the radii of very massive stars (with masses $2~\Ms$).
Both statements imply that $f_2$ is mainly sensitive to the high-density EoS. 
Combined with observations of the binary neutron star inspiral, these generalized quasi-universal
relations can be used to simultaneously infer the characteristic radius
and slope of the neutron star mass-radius relation.
\end{abstract}


 \section{Introduction}

The advent of gravitational wave (GW) astronomy has opened an exciting new
era for constraining the equation of state (EoS) of ultra-dense matter. Observations 
of the inspiral from the first binary neutron star merger, GW170817 \citep{TheLIGOScientific:2017qsa}, have already 
provided strong constraints on the EoS (see e.g.,
\citealt{Baiotti:2019sew,Raithel:2019uzi,GuerraChaves:2019foa,Chatziioannou:2020pqz} for recent reviews),
 and it is expected that the future detections of post-merger GWs,
 which will become possible with improved sensitivity of the detectors  (e.g., \citealt{Torres-Rivas:2018svp}),
 will offer further insight \citep{Baiotti:2016qnr,Paschalidis:2016vmz,Bauswein:2019ybt,Bernuzzi:2020tgt,Radice:2020ddv}. 

 The post-merger GWs are emitted by oscillations of the hot,
rapidly rotating, massive neutron star remnant following the merger. Numerical
simulations have shown that the spectra of these post-merger GWs display distinct peaks, which
 are driven by various oscillation modes of the remnant and thus encode information about its stellar structure.
For example, the dominant spectral peak, which we call $f_2$
and which is present in essentially all numerical simulations of the post-merger phase,
is powered by quadrupolar oscillations of the remnant(e.g.,
\citealt{Stergioulas:2011gd,Takami:2014tva,Rezzolla:2016nxn}). 
Many studies have shown that this $f_2$ spectral
peak correlates strongly with the characteristic radius, $R$, of the neutron star EoS (e.g., \citealt{Bauswein:2012ya,Bauswein:2011tp,Takami:2014zpa,Bernuzzi:2015rla}).
In one recent meta-analysis of over 100 numerical simulations, \citet{Vretinaris:2019spn}
reported a latest set of quasi-universal relations between $f_2$ and the radius at various masses,
concluding that the correlation was strongest for the radius of a 1.6~$\Ms$ star.
Such quasi-universal relations provide a straightforward way of mapping the post-merger GWs to the EoS, 
either by direct comparison against X-ray measurements of the neutron star radius \citep[e.g.,][]{Ozel:2016oaf,Miller:2021qha,Riley:2021pdl} 
or by enfolding the $R(f_2)$ constraint into the Bayesian inference schemes that have been 
 developed to constrain the EoS from mass/radius observations \citep[e.g.,][]{Ozel:2015fia,Steiner:2015aea,Raithel:2017ity,Raaijmakers:2021uju}.
As such, these quasi-universal relations are a powerful and important tool for accurately interpreting the
upcoming detections of post-merger GWs from a binary neutron star coalescence.

In this Letter, we report on new violations of the quasi-universal relations between $f_2$ and $R$, using
a diverse set of EoS models which are systematically constructed to span a wide range of slopes in their
mass-radius $\left(M-R\right)$ relations. It has previously been shown that the quasi-universal relations break down for EoSs with a strong, first-order phase transition \citep{Bauswein:2018bma}, which leads to much smaller radii at high neutron star masses.
In this Letter, we generalize this result and demonstrate that the standard quasi-universal relations generically break down for EoSs 
that predict a significantly non-vertical mass-radius slope. 
The violations are the most extreme for models that predict increasing
 radii at high masses (corresponding to a stiffening in the EoS at high densities). 
We find strong evidence that the quasi-universal relations need to be generalized to include the slope
of the mass-radius curve as an additional parameter. Alternatively, quasi-universality can also be restored in our sample
by correlating $f_2$ with the radius at higher masses than are typically considered ($M\sim2~\Ms$).

These findings imply that the post-merger GWs are mainly sensitive to the EoS at high densities.
Combined with observations of the inspiral (which are goverened by lower-density physics),
a complete GW event could thus constrain the EoS across a potentially wide range of densities. In more 
concrete terms, our findings suggest that a measurement of $f_2$, if supplemented with additional
information from the inspiral, can be used to constrain not
only the characteristic radius of a neutron star, but also the slope of the mass-radius relation.

This Letter is laid out as follows. In Sec. \ref{sec:methods} we provide an overview of the microphysics and numerical methods used in this work. In Sec. \ref{sec:results} we present the results of our numerical simulations, and provide a detailed analysis of the GW peak frequencies. In Sec. \ref{sec:discussion} we discuss our findings in the context of other observations of the neutron star radius.

\section{Methods}\label{sec:methods}
In the following, we will provide a quick summary of the EoS models and numerical methods employed in this work. 
For ten of the fourteen models considered in this work, we describe the nuclear composition of the neutron stars using a finite-temperature EoS framework laid out in \citet{Raithel:2019gws} (for additional details on our implementation, see \citealt{Most:2021ktk}). In particular, 
the zero-temperature, $\beta$-equilibrium pressure is constructed using a piecewise polytropic 
parametrization with five segments \citep{Ozel:2009da, Read:2008iy, Steiner:2010fz, Raithel:2016bux}.
We extrapolate this parametric, cold EoS to finite temperatures using the $M^*$-model of \citet{Raithel:2019gws}, which includes the leading-order effects of degeneracy 
 in the thermal pressure. We additionally extrapolate the EoS to arbitrary proton fractions using an 
 approximation of the nuclear symmetry energy \citep{Raithel:2019gws}. 
The low-density EoS (below 0.5 times the nuclear saturation density) is taken to be the finite-temperature SFHo EoS
 \citep{Steiner:2012rk},\footnote{The SFHo table was provided by \texttt{stellarcollapse.org}.} 
which we smoothly connect to our high-density parametrization using the free-energy matching procedure of 
\citet{Schneider:2017tfi} to ensure that the resulting EoS is thermodynamically consistent. 
In constructing these parametric EoSs, we use the same $M^*$-parameters as in  
\citet{Most:2021ktk}: namely, $n_0=0.12$~fm$^{-3}$, $\alpha=0.8$, and $\gamma=0.6$, which are consistent
with best-fit parameters for a range of commonly used tabulated EoSs \citep{Raithel:2019gws}. For
additional details on the construction of our, see \citet{Most:2021ktk}.

\begin{figure}[!ht]
\centering
\includegraphics[width=0.5\textwidth]{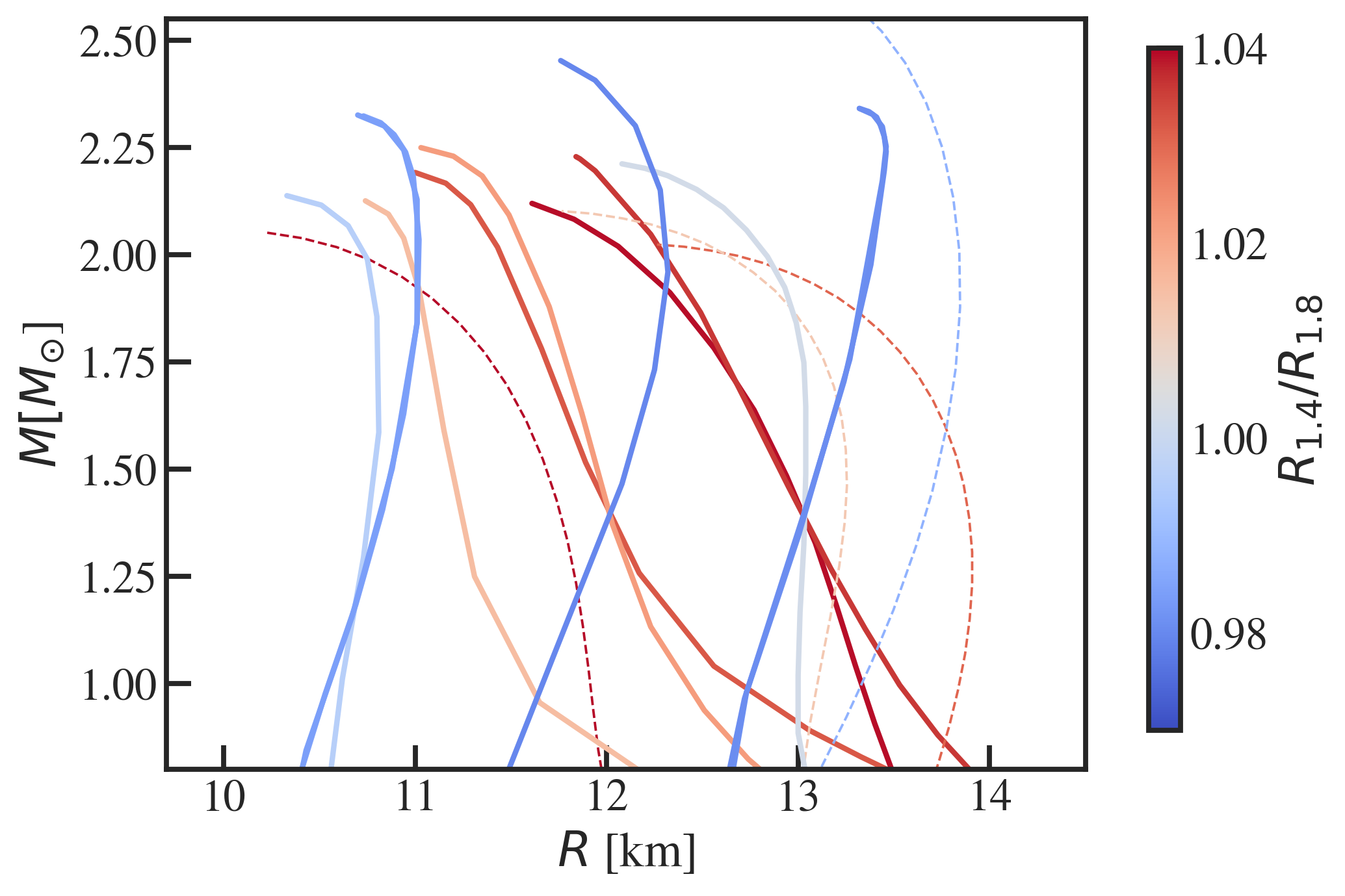}
\caption{ \label{fig:EoS} Mass-radius relations for the equations of state simulated in this work.
The solid lines correspond to our parametric models, while the dashed line represent four commonly used 
finite-temperature EoS tables included in this work (in order of increasing radius, these are: SLy4,
BHB$\Lambda\phi$, TMA for the red curves, and LS375 in blue).
 We color-code the curves according to $R_{1.4}/R_{1.8}$, where $R_x$ represents the radius of
a star with mass $M=x\Ms$.  The parametric EoS sample (solid lines) is specifically constructed
to sample a wide range of slopes
in the mass-radius relations.}
\end{figure}

In total we construct ten EoSs for this study, seven of which have been presented before \citep{Most:2021ktk}. We show their mass-radius relations in Fig.~\ref{fig:EoS} , color-coded throughout this work according to the parameter $R_{1.4}/R_{1.8}$, where $R_x$
 indicates the radius of a neutron star of mass $M=x\Ms$. 
 The three new models constructed for this work correspond to 
 the two blue curves with $R_{1.4}=13$~km, as well as the darker blue curve with $R_{1.4}=10.8$~km.  We provide additional details on the three new EoS models in Appendix~A.

Different from previous works (e.g.,\citealt{Takami:2014zpa,Vretinaris:2019spn}), our EoS sample spans a wide range of slopes in the mass-radius relation, with multiple EoSs that predict a backwards-bending mass-radius slopes. In particular, we construct families of EoSs that have an identical characteristic radius $R_{1.4}$, but that vary significantly in their radii at higher masses. This allows us, for the first time, to systematically investigate how the common set of quasi-universal relations for the post-merger GW frequency $f_2$ break down when the sample of EoS is significantly broadened. \\

In addition to these parametrically-constructed EoSs, we also simulate four existing, finite-temperature EoS tables:
the nuclear EoS of 
\citet{Lattimer:1991nc} 
with a bulk incompressibility of 375~MeV (``LS375");
the nuclear EoS SLy4 
\citep{Schneider:2017tfi}
; the nuclear EoS TMA 
\citep{Toki:1995ya,Hempel:2009mc}
, and the hyperonic EoS BHB$\Lambda\phi$ 
\citep{Banik:2014qja}
. With this comprehensive EoS sample, we can assess the validity of our findings for both commonly used and newly constructed  EoS tables.

In  order to investigate the impact of the different mass-radius slopes on the post-merger GW emission,
we simulate the coalescence of a GW170817-like event for each of the fourteen EoSs in our sample. For our baseline simulations, we consider a moderate mass ratio of $q=0.85$ for a
system with a total mass of $M=2.73\, M_\odot$. For a subset of models, we also perform simulations with the same chirp mass, but with $q=1$. For each EoS, we construct numerical initial conditions of
compact binaries on quasi-circular orbits using the \texttt{LORENE} code \citep{Gourgoulhon:2000nn}.
We then use the \texttt{Frankfurt-/IllinoisGRMHD (FIL)} code \citep{Most:2019kfe,Etienne:2015cea} to solve the coupled
Einstein-hydrodynamics system \citep{Duez:2005sf} using the Z4c formulation \citep{Hilditch:2012fp}.
\texttt{FIL} operates on top of the Einstein Toolkit infrastructure \citep{Loffler:2011ay,Schnetter:2003rb}.
A detailed description of the numerical setup can be found in \citet{Most:2021ktk}.

\section{Results}\label{sec:results}

 \begin{figure*}[!ht]
\centering
 \includegraphics[width=0.8\textwidth]{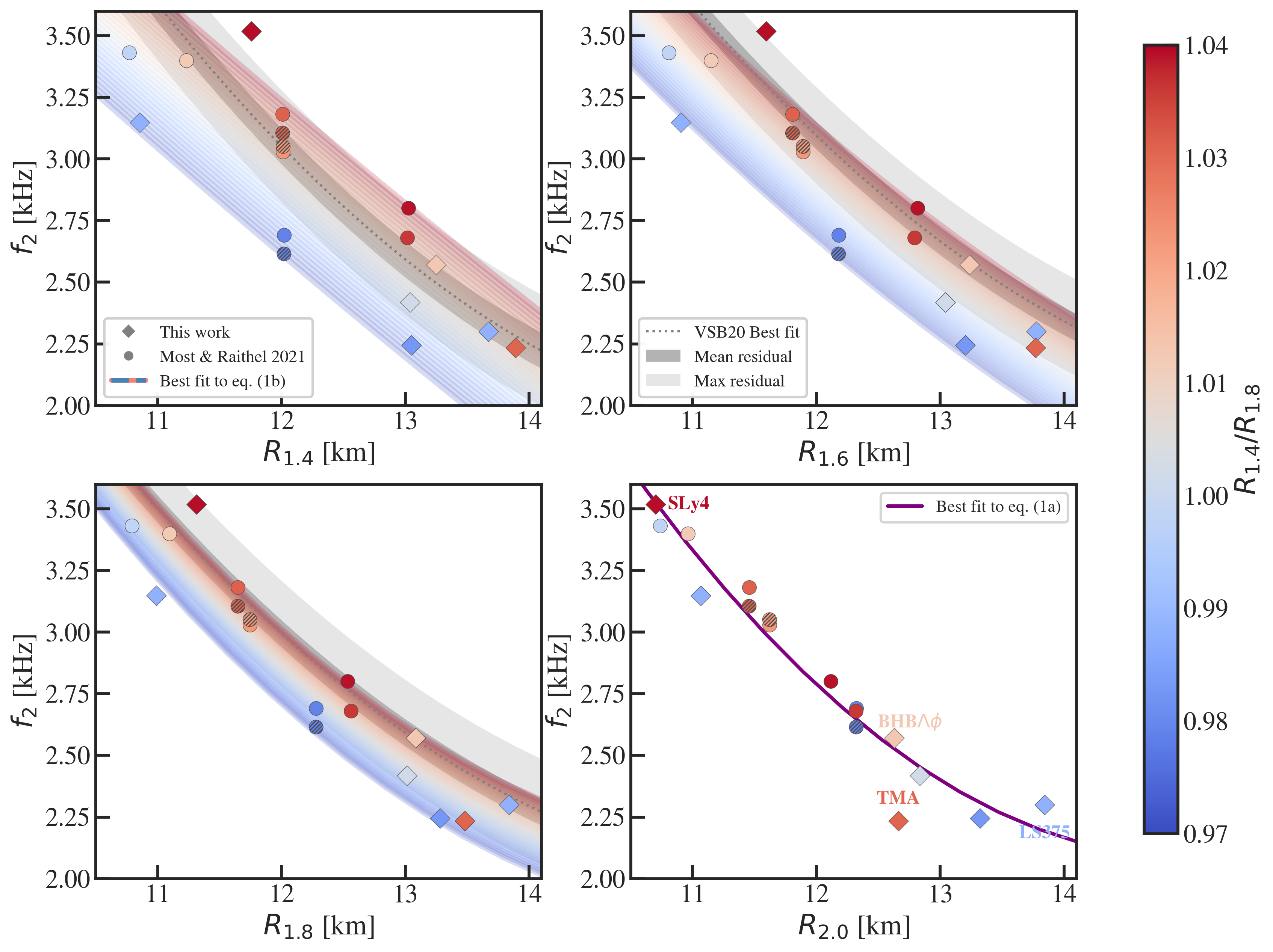}
 \caption{\label{fig:corr} Peak frequency of the post-merger GW spectrum as a function of the neutron star radius at various masses. The color
coding is the same as in Fig.~\ref{fig:EoS}. Diamonds represent results from the new simulations
performed in this work, while circles correspond to results from \citet{Most:2021ktk}. We note that the results
from the four microphysical EoS tables are included in all four panels, but are only labeled in 
the bottom right panel for clarity. Solid-filled
symbols indicate a binary mass ratio of $q=0.85$, while the hatched symbols are for $q=1$. The dotted gray
lines correspond to the best-fit, quasi-universal relations reported in \citet{Vretinaris:2019spn},
with mean and maximum residuals from that work shown in the dark and light gray bands, respectively.
Finally, the blue and red lines correspond to our best-fit, two-parameter quasi-universal relations
(eq.~\ref{eq:two_param}), with varying values of  $R_{1.4}/R_{1.8}$; while the purple line corresponds to the best-fit, single-parameter relation for $f_2(R_{2.0})$.
We find that models with $R_{1.4}/R_{1.8}<1$ (blue points) systematically violate the existing
quasi-universal relations for stars of intermediate mass (top row) and instead follow a separate relation between $f_2$ and $R_x$.}
\end{figure*}

In this section, we present the results of our numerical simulations.
We focus exclusively on the quasi-universal behavior of the post-merger GW frequency spectrum.
These are computed as outlined in Appendix~C of \citet{Most:2021ktk}.
From these spectra, we identify $f_2$ as the frequency corresponding to the maximum power. 
We show $f_2$ as a function of the radius at various masses in Fig.~\ref{fig:corr}, for
each of the EoSs simulated in this work, as well as a previous set of simulations from \citet{Most:2021ktk}.

Figure~\ref{fig:corr} also shows, for reference, the quasi-universal relations found in \citet{Vretinaris:2019spn} between
$f_2$ and $R_{1.4}$, $R_{1.6}$, and $R_{1.8}$. Correlations with $R_{2.0}$ were not studied in that work.
The dashed gray line shows the best fit relations from \citet{Vretinaris:2019spn}, while the dark 
and light gray bands represent the mean and maximum residuals, respectively. 
Overall, we find a similar inverse correlation between $f_2$ and the radius at any mass, but one
that is much broader than reported in \citet{Vretinaris:2019spn}. In particular, we find that a large number of our models violate the existing quasi-universal relations. For example, when considering the correlations with $R_{1.4}$
or $R_{1.6}$ (top row of Fig.~\ref{fig:corr}), we find that 5-6 of the models fall outside of the previous maximum-residual error
bands, and a majority fall outside of the mean-residual error band of \citet{Vretinaris:2019spn}.
This violation of the quasi-universal relations is extreme, with $f_2$ varying by up to $\sim\,$600~Hz for models with the same
 characteristic $R_{1.4}$. The scatter is largest for the correlation with $R_{1.4}$, and decreases at increasing masses. 
Only when correlating $f_2$ as a function of $R_{1.8}$ do we find that most of our models fall within the previous maximum error
band. However, even in this case, many of the EoSs populate the extreme lower edge of that
error band.

In all three of these cases ($R_{1.4},\ R_{1.6}\,\ \text{and}\  R_{1.8}$), we find a strong trend between the degree to which the 
quasi-universal relations are violated and the slope of the mass-radius curve, which is here characterized by the ratio $R_{1.4}/R_{1.8}$ (for alternate definitions of the slope, see the discussion in the Appendix~B). The models
that most strongly violate the existing quasi-universal relations are those with $R_{1.4}/R_{1.8} < 1$,
which corresponds to the EoSs with a backwards-bending mass-radius relation in Fig.~\ref{fig:EoS}.
This phenomenological $M-R$ behavior is caused by a stiffening of the EoS at high densities, 
 such that more massive stars are characterized by steeper pressure gradients, and thus extend to larger radii.
This type of behavior is consistent with the predictions, for example, of the emergence of a quarkyonic phase 
 of matter \citep{McLerran:2018hbz}. On the other hand, an EoS with significant softening at high densities  (caused, e.g., by a cross-over phase transition) 
 is characterized by a forwards-tilting $M-R$ relation, with decreasing radii at high masses.
In general, we find that models with $R_{1.4}/R_{1.8}<1$ (EoSs with significant stiffening at high densities),
lead to systematically lower values of $f_2$, while models with $R_{1.4}/R_{1.8} > 1$ 
(EoSs with significant softening), lead to larger $f_2$. Indeed, the red and blue points in Fig.~\ref{fig:corr} appear to follow separate quasi-universal relationships between $f_2$ and $R_x$. 

We note that the correlations for EoSs with forward-bending $M-R$ curves (red points) are
still largely consistent with the existing quasi-universal relations.
In the limit of extreme softening in the EoS (i.e., a first-order phase transition to a stable hybrid star), \cite{Bauswein:2018bma}
have shown that the $f_2$ quasi-universal relations can also be broken, and that this feature
can be used to infer the presence of a phase transition from the post-merger GWs. The trend reported in that work
is the same as we find in Fig.~\ref{fig:corr} -- i.e., that EoSs with softening tend to have larger $f_2$ -- but
our family of EoSs with softening are less extreme (i.e., no first-order phase transitions) and thus show more modest violations in that direction. 

We confirm that these trends between $f_2$ and the mass-radius slope are not very sensitive
to the mass ratio of the binary by performing a subset of the simulations at a second mass ratio
of $q=1$. These points (which correspond to the three models with $R_{1.4}=12$~km)
are shown in Fig.~\ref{fig:corr} with hatched shading. We find small shifts in $f_2$ depending on the mass ratio,
as has also been found in previous studies (e.g., \citealt{Bernuzzi:2015rla,Rezzolla:2016nxn}), but that the overall trend with $R_{1.4}/R_{1.8}$ is maintained.

In order to confirm that our results are robust to the parametric construction of the EoSs,
we perform one further simulation with an approximate version of the TMA EoS, where the finite-temperature
and composition-dependent dimensions of the EoS table have been replaced by the complete $M^*$-framework,
using the TMA-specific parameters listed in \citet{Raithel:2019gws}.
The peak frequency extracted with the parametric version of the EoS table is consistent
with the $f_2$ obtained using the full table to within 20~Hz, thus confirming that the parametric thermal treatment
produces reliable peak frequencies. A further comparison of $M^*$-parametric EoS
to tabulated EoS in the context of merger simulations is presented in \cite{Raithel:2022nab}.

In addition, we find that the three microphysical EoS tables, SLy4, BHB$\Lambda\phi$, and LS375, all support the trend 
we have identified with the parametric EoSs. In particular, LS375 leads to a smaller-than-expected $f_2$,
while SLy4 leads to a larger $f_2$ than predicted by the previous quasi-universal relations. Both
of these deviations are explained by the mass-radius slopes of these EoSs. 
The peak frequency for BHB$\Lambda\phi$, which predicts only a small MR-slope, lies 
closer to the existing quasi-universal relation, also as expected. The peak frequency for TMA (the right-most red diamond in Fig.~\ref{fig:corr}) is a modest outlier,
which we attribute to the changing curvature of its mass-radius slope. We discuss this point further in Appendix~B.

In order to quantify these trends with mass-radius slope, we fit the results shown 
in Fig.~\ref{fig:corr} to two different functional forms,
using a standard, non-linear, least-squares fitting algorithm.
The first functional form is a single-parameter relation motivated by the fits done in \citet{Vretinaris:2019spn}, 
which were quadratic in $R_x$ (where $x$ represents an arbitrary stellar mass). We also consider a two-parameter
relation, which adds a linear correction term that scales with $R_{1.4}/R_{1.8}$, in order to account
for the trends observed in Fig.~\ref{fig:corr}.\footnote{We point out that to linear order, 
\begin{align*}
    \frac{R_{1.4}}{R_{1.8}} \approx 1 - \frac{2}{5\, R_{1.8}} \left. \frac{{\rm d} R}{{\rm d} M}\right|_{M=1.8\,M_\odot}\,.
\end{align*}
Hence, a linear correction in $\frac{R_{1.4}}{R_{1.8}}$ is equivalent to a linear correction in the slope $\frac{{\rm d} M}{{\rm d} R}$, with an adjusted set of fit coefficients. We report the corresponding fits for alternate definitions of the slope parameter in Appendix~B.} 
The fitting functions are thus 
\begin{subequations}
\label{eq:quad}
\begin{align}
\label{eq:one_param}
f_2(R_x) &= b_0 + b_1 R_x  + b_2 R_x^2  \\
\label{eq:two_param}
f_2(R_x,\frac{R_{1.4}}{ R_{1.8}}) &= b_0 + b_1 R_x + b_2 R_x^2 + b_3 \left( \frac{R_{1.4}}{ R_{1.8}} \right),
\end{align}
\end{subequations}
where $b_i~(i \in [0,3])$ are the fit coefficients.
We note that we do not include the chirp mass as a free parameter (in contrast to \citealt{Vretinaris:2019spn}) because
all simulations done in this work are for a fixed $\mathcal{M}_c=1.186~\Ms$. Thus, our fits can be viewed as 
characterizing one slice of the $f_2 - R_x - \mathcal{M}_c$ plane that was identified in \citet{Vretinaris:2019spn}.
The best-fit coefficients for eq.~(\ref{eq:quad}) are reported in Table~\ref{table:fits}, for various choices of $R_x$. To illustrate the performance of these fits, Fig.~\ref{fig:corr} additionally shows the best-fit, two-parameter relations for $R_{1.4}$\,,$R_{1.6}$ and $R_{1.8}$ with varying choices of the slope parameter (shown in the light blue-to-red shading), as well as the best-fit, single-parameter relation for $R_{2.0}$ (in purple).
Finally, Table~\ref{table:fits} also reports the adjusted coefficient of determination ($\mathcal{R}^2$) for each fit,  the Bayesian Information Criteria (BIC),\footnote{In order to be as conservative as possible, 
  we use the BIC, rather than other criteria such as the Aikake Information Criteria,
  as the BIC more strongly penalizes the addition of free parameters to the model. We compute the BIC under the 
  assumption that the errors in $f_2$ are independent , identical, and Gaussian. 
  When comparing two models, $\Delta$BIC$>$5 indicates ``strong" evidence and $\Delta$BIC$>$10 indicates ``decisive"
   evidence in favor of the model with a more negative BIC, according to the Jeffreys
  scale \citep{Jeffreys61,Liddle:2007fy}.}
   and the maximum and mean residuals.

\centerwidetable
\begin{deluxetable*}{llllllllll}
\tabletypesize{\footnotesize}
\tablewidth{0.6\textwidth} 
\tablecaption{\label{table:fits} Fit coefficients for eqs.~(\ref{eq:one_param}) and (\ref{eq:two_param}). All coefficients assume radii in km and $f_2$ in kHz. The right-most four
columns contain the adjusted coefficient of determination ($\mathcal{R}^2$), the Bayesian information criterion, the maximum residual,
and the mean residual for each fit. In this table, $R_x$ indicates the radius of a neutron star with mass $M=x\Ms$, while $R_{\rm max}$ indicates the radius corresponding to the maximum mass configuration.} 
\tablehead{
\colhead{ $R_x$ }  &
\colhead{$b_0$  }  &
\colhead{$b_1$ }  &
\colhead{$b_2$}  &
\colhead{$b_3$}  &
\colhead{Adjusted $\mathcal{R}^2$} &
\colhead{BIC }  &
\colhead{Max resid (kHz)}  &
\colhead{Mean resid (kHz)}
}
\startdata   
 $R_{1.4}$   & 1.754  & 0.573  & -0.039  & $-$  & 0.680  & 3.5  & 0.43 & 0.17\T\\
 $R_{1.6}$   & 5.384  & 0.001  & -0.017  & $-$  & 0.793  & -3.9  & 0.38 & 0.13\\
 $R_{1.8}$   & 11.379  & -0.964  & 0.022  & $-$  & 0.883  & -13.5  & 0.29 & 0.10\\
 $R_{2.0}$   & 19.554  & -2.328  & 0.078  & $-$  & 0.932  & -22.9  & 0.28 & 0.07\\
 $R_{\rm max}$   & 21.740  & -2.833  & 0.102  & $-$  & 0.919  & -19.9  & 0.22 & 0.09\B\\
 \hline
 $R_{1.4}$   & 0.081  & -0.592  & 0.007  & 8.900  & 0.927  & -20.2  & 0.21 & 0.08\T\\
 $R_{1.6}$   & 5.403  & -1.083  & 0.027  & 6.544  & 0.923  & -19.3  & 0.21 & 0.08\\
 $R_{1.8}$   & 12.179  & -1.821  & 0.057  & 4.287  & 0.933  & -21.7  & 0.20 & 0.07\\
 $R_{2.0}$   & 19.455  & -2.459  & 0.083  & 0.837  & 0.929  & -20.6  & 0.30 & 0.07\\
 $R_{\rm max}$   & 21.421  & -2.871  & 0.104  & 0.504  & 0.913  & -17.2  & 0.23 & 0.08\B\\
 \enddata
\end{deluxetable*}

We turn first to the results of the single-parameter fits. The strength of the 
correlation between $f_2$ and $R_x$ increases significantly as we consider larger masses $x$, 
as expected based on the scatter seen in Fig.~\ref{fig:corr}. For example, the adjusted $\mathcal{R}^2$ for
the single-parameter fit to $R_{1.4}$ is only 0.68, but it increases to 0.93 for $R_{2.0}$.  The $\Delta$BIC also
shows strong evidence for each subsequently larger mass compared to the previous 
(e.g., $R_{1.8}$ is favored over $R_{1.6}$, etc.). The correlation is strongest for $R_{2.0}$.
If we correlate instead with the radius corresponding to the maximum mass (as in, e.g., \citealt{Bauswein:2011tp}),
the adjusted $\mathcal{R}^2$ decreases and the correlation is disfavored, compared to $R_{2.0}$.

When comparing the results of eq.~(\ref{eq:one_param}) and (\ref{eq:two_param})  with our sample of EoSs, 
we find strong evidence in favor of adding this second parameter 
to the existing quasi-universal relations. For example, when comparing the evidence for a single-parameter
fit with $R_{1.4}$ to the two-parameter fit with $R_{1.4}/R_{1.8}$, we find $\Delta$BIC$\approx24$,
indicating decisive evidence for the latter model. The results are similar for $R_x=R_{1.6}$ 
(``decisive" evidence with $\Delta$BIC=15) and $R_{1.8}$ (``strong" evidence with $\Delta$BIC=8).
 For the case of $R_x = R_{2.0}$, the
correlation with the single-parameter fitting formula is already quite strong, and we do not
find statistical evidence to justify the addition of a second parameter.

To summarize, we find that the existing, single-parameter quasi-universal relations all break down for 
EoSs with significant stiffening at high densities, i.e., with backwards-bending mass-radius relations.
We find that for $R_{1.4}$, $R_{1.6}$, and $R_{1.8}$, the existing single-parameter relations are all significantly
 improved with the addition of a second parameter, which incorporates information about the $M-R$ slope.
 Indeed, by expanding the functional forms to include a second parameter ($R_{1.4}/R_{1.8}$), we are able to
 recover quasi-universality, as evidenced by coefficients of determination near unity. 
 Alternatively, we are also able to better maintain quasi-universality for the correlation between $f_2$ and a single radius
 at sufficiently high masses, here for $R_{2.0}$.

 The fact that quasi-universality is recovered for radii at very high masses strongly suggests that $f_2$ depends 
 on the EoS at high densities. In addition, the slope of the $M-R$ relation has been shown to correlate 
 with the pressure at $\sim3.7 \rho_{\rm sat}$ \citep{Ozel:2009da}, where $\rho_{\rm sat}\approx2.7\times10^{14}$~g/cm$^{3}$  is 
 the nuclear saturation density. Thus, by adding a slope-dependent parameter to the
 fitting function for $R_x$ at low masses, we are effectively adding in information about the
 high-density EoS, which also helps to restore the quasi-universality.

We note that for EoSs that predict mass-radius curves with significant curvature in
their slopes, such as TMA, the choice of where to define a mass-radius slope is not always obvious.
In this work, we use a definition of the slope that leads to the best goodness-of-fit statistics.
We discuss alternate definitions of the slope and their (negligible) impact on our conclusions
in  Appendix~B.

 \begin{figure}[!ht]
\centering
\includegraphics[width=0.5\textwidth]{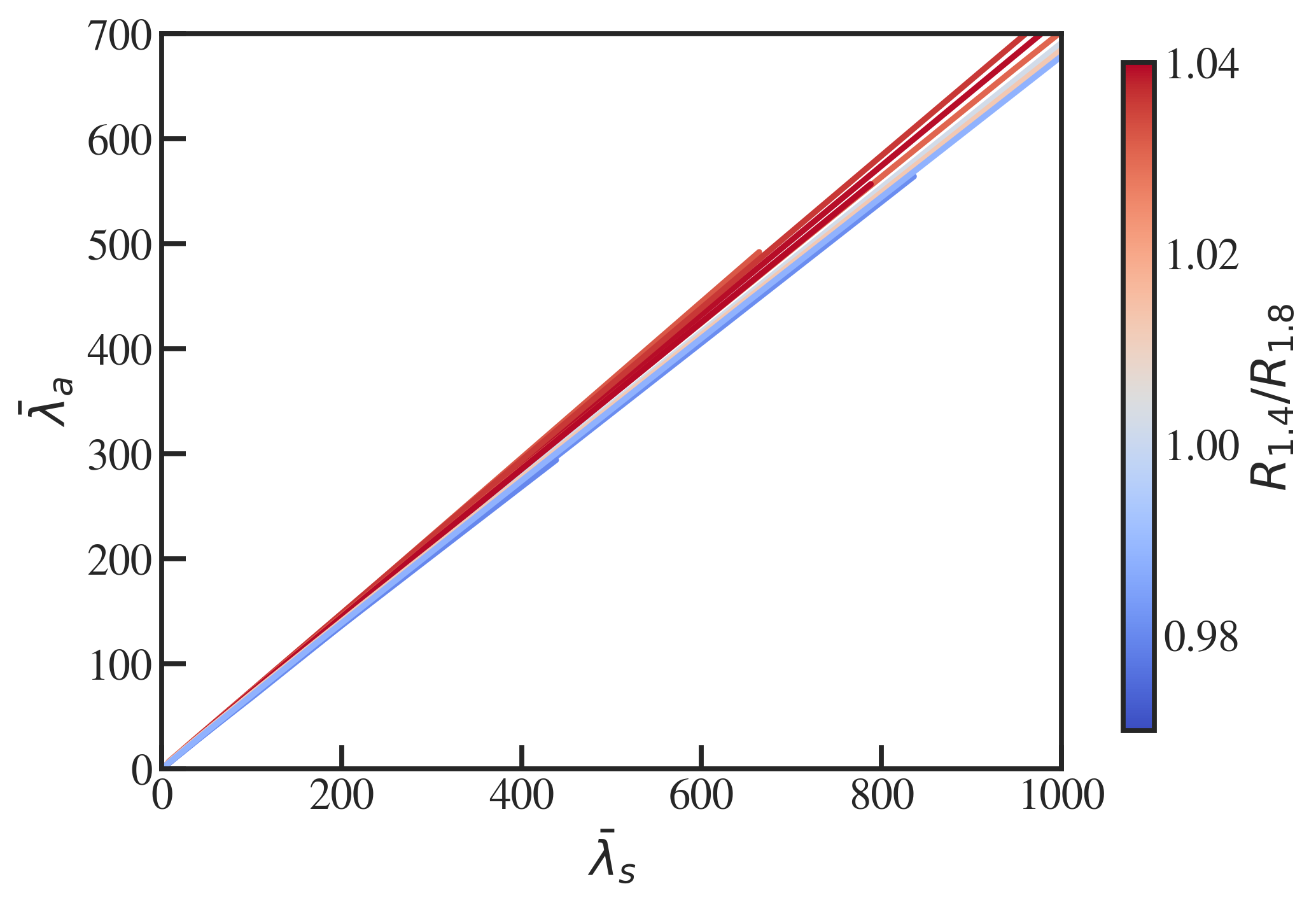}
\caption{\label{fig:binary_love} 
Binary Love relations for the symmetric 
$\bar{\lambda}_s$ and antisymmetric, $\bar{\lambda}_a$ tidal deformabilities of the coalescing binary
with mass ratio $q=0.85$ considered in this work. Different curves correspond to the various
equation of state models used, with color indicating the slope of the mass-radius curve in terms 
of the ratio $R_{1.4}/R_{1.8}$, where $R_x$ denotes the radius of a neutron star with mass $M= x\,M_\odot$.
}
\end{figure}

\section{Discussion and Conclusions} \label{sec:discussion}

In this Letter, we have shown that the existing family of single-parameter,
quasi-universal relations are insufficient to reliably infer the
radius of an intermediate-mass neutron star from the
post-merger GW frequencies alone.  EoSs
that produce backwards-bending mass-radius curves lead to systematically
lower values of the post-merger peak frequencies $f_2$. As a result, we show
that such EoSs violate existing quasi-universal relations proposed to infer
neutron star radii, e.g., $R_{1.4}$, $R_{1.6}$, or $R_{1.8}$, from the
post-merger frequency $f_2$. Moreover, we find that the EoSs that violate the
existing relations form their own, separate quasi-universality class,
which can be used to relate $f_2$ to neutron star radii {for a fixed mass-radius slope} (see Fig. \ref{fig:corr}).
This observation motivates the extension of existing quasi-universal relations to
incorporate a second parameter, related to the slope of the mass-radius
curve. For the sample of EoSs explored in this Letter, we find strong statistical evidence in favor of the two-parameter,
quasi-universal relations, compared to models that do not incorporate information
about the mass-radius slope.

Interestingly, we find that single-parameter quasi-universality can be
better maintained when considering massive neutron stars $M\simeq 2\, M_\odot$.
Intuitively, this is consistent with the fact that the post-merger
remnant itself is a massive neutron star, and thus probes higher densities than are
present in the inspiral.  Hence, we might reasonably expect that the peak frequencies
of the post-merger GWs should correlate more strongly with parameters of the high-density
EoS (this is also consistent with our findings in \citealt{Most:2021ktk}).

The better universality of the $f_2-R_{2.0}$ relation is of particular interest,
given the recent NICER observations of 
the $\sim2~\Ms$ neutron star PSR J0740+6620, for which two recent radius inferences have been
performed \citep{Miller:2021qha,Riley:2021pdl}.
Based on our findings,
we expect that a future measurement of $f_2$ will be be able to provide robust and
independent constraints on the radius of this pulsar.

If such a measurement of $R_{2.0}(f_2)$ can be supplemented with
constraints on $R_{1.4}$ from the inspiral measurement of the tidal deformability 
(see, e.g., \citealt{Baiotti:2019sew,Raithel:2019uzi,Chatziioannou:2020pqz}), then a sufficiently sensitive merger event -- observed
from inspiral through postmerger -- could effectively
be used to the trace out the \textit{entire} mass-radius 
relation, to linear order.

Alternatively, it may also be possible to reconstruct the linearized mass-radius curve
by utilizing the so-called binary Love relations of \citet{Yagi:2015pkc}. 
These relations relate symmetric, 
$\bar{\lambda}_s = \frac{1}{2}\left( \bar{\lambda}_1 + \bar{\lambda}_2\right)$, 
and antisymmetric, $\bar{\lambda}_a = \frac{1}{2}\left( \bar{\lambda}_1 - \bar{\lambda}_2\right)$, 
combinations of the mass-normalized tidal deformabilities $\bar{\lambda}_{1,2}$ of the
inspiralling neutron stars in an EoS-insensitive way \citep{Yagi:2015pkc}.
In a recent work, \citet{Tan:2021nat} showed that these relations,
previously thought to be fully universal, also receive an effective
correction that is linearly proportional to the slope of the mass-radius curve.
We illustrate this behavior in Fig. \ref{fig:binary_love}, where we
show the $\bar{\lambda}_a-\bar{\lambda}_s$ correlation for the EoS and binary
parameters used in this work. We can clearly see that, in line with
\citet{Tan:2021nat}, there is a range of slopes separating EoSs with backwards- and
forwards-bending mass-radius curves. Because the binary-Love relation also depends
on the neutron star radius and mass-radius slope, but with a different dependence
than our two-parameter, quasi-universal relations for $f_2$,
combined detections of inspiral and post-merger should enable a reconstruction of
the linearized mass-radius curve. 

While our findings constitute strong evidence for the existence of a two-parameter, 
quasi-universal relation for $f_2(R_x, R_{1.4}/R_{1.8}$) in our sample,
future work will be necessary to further quantify this new dependency on the mass-radius slope.
In particular, this will require a systematic investigation of an even larger number of EoSs with 
varying and non-linear mass-radius slopes, and simulations of a wide class of binary masses and mass ratios.
We leave such detailed explorations to future work.
\\

\section*{Acknowledgements}

ERM thanks J. Noronha-Hostler and N. Yunes for insightful discussions related to this work. 
CAR and ERM gratefully acknowledge support from postdoctoral fellowships
at the Princeton Center for Theoretical Science, the Princeton Gravity
Initiative and the Institute for Advanced Study. CAR is additionally supported
as a John N. Bahcall Fellow at the Institute for Advanced Study.
Part of the simulations presented in this article were performed on computational resources managed and supported by Princeton Research Computing, a consortium of groups including the Princeton Institute for Computational Science and Engineering (PICSciE) and the Office of Information Technology's High Performance Computing Center and Visualization Laboratory at Princeton University.
This work also used the Extreme Science and Engineering Discovery Environment (XSEDE) through Expanse at SDSC and Bridges-2 at PSC, via allocations PHY210053 and PHY210074. 
The authors also acknowledge the Texas Advanced Computing Center (TACC) at The University of Texas at Austin for providing HPC resources that have contributed to the research results reported within this paper, under LRAC grants AT21006.

\software{Einstein Toolkit \citep{Loffler:2011ay},
          \texttt{Carpet} \citep{Schnetter:2003rb},
          \texttt{Frankfurt-/IllinoisGRMHD (FIL)} \citep{Most:2019kfe,
	  Etienne:2015cea},
	  \texttt{LORENE} (\url{https://lorene.obspm.fr}), 
	  Matplotlib \citep{Hunter:2007},
	  seaborn \citep{Waskom2021}
          }

\appendix

\section{New piecewise polytropic EoS models studied in this work}

\begin{figure*}[!ht]
\centering
\includegraphics[width=0.5\textwidth]{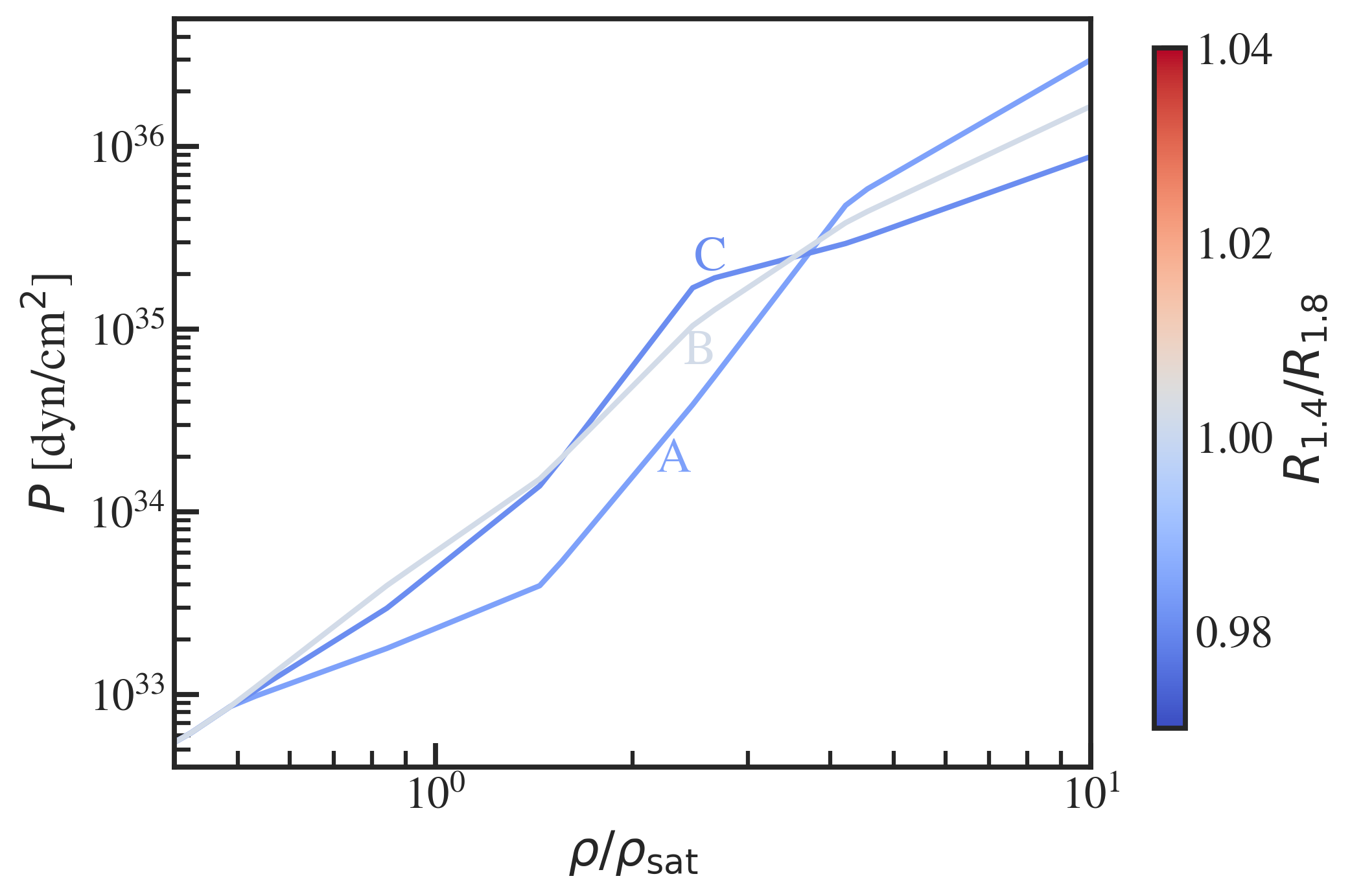}
\caption{\label{fig:newEOS} Pressure as a function of density for the three piecewise-polytropic EoSs newly constructed in this work. The density, $\rho$, is plotted with respect to the nuclear saturation density, $\rho_{\rm sat}=2.7\times10^{14}\text{g/cm}^3$. The labels are arbitrarily assigned, for reference in Table~\ref{table:EOS_params}.}
\end{figure*}

In this appendix, we provide additional details on the new piecewise polytropic EoSs
constructed for this work. As described in Sec.~\ref{sec:methods}, seven of the ten piecewise polytropic EoSs studied in this work were previously presented in \cite{Most:2021ktk}, and we refer the reader to that paper for details on those models. We show the three new models, which were constructed specifically for this study, in Fig.~\ref{fig:newEOS}. These models correspond to some of the most extreme backwards-bending mass-radius relations from Fig.~\ref{fig:EoS}. As can be seen in Fig.~\ref{fig:newEOS}, the backwards-bending phenomenology is produced by a stiffening of the EoS at intermediate densities.

As was the case for the original sample of seven EoSs constructed in \citet{Most:2021ktk}, these three new models are subject to several physical constraints. In particular, we require that each of these EoSs remain causal and thermodynamically stable, and that the maximum mass of each EoS is at least 2~$\Ms$, in order to be consistent with the observation of massive pulsars \citep{Demorest:2010bx,Antoniadis:2013pzd,Fonseca:2016tux,NANOGrav:2019jur}. In addition, we impose a lower limit on the pressure at our first fiducial density $P(0.86~\rho_{\rm sat}) \gtrsim 1.7 \times 10^{33} \text{dyn/cm}^2$, which is set by the two-body potential of Argonne AV8 \citep{Gandolfi:2013baa}.

The radii and tidal deformabilities of these EoS models are presented in Table~\ref{table:EOS_params}. In addition, Table~\ref{table:EOS_params} also reports the piecewise polytropic parameters that describe each EoS. We note that all of the phenomenological EoS models used in this work comprise five piecewise polytropic segments, which are spaced log-uniformly in the density. The fixed dividing densities are located at [0.86, 1.47, 2.52, 4.32, and 7.4] $\times \rho_{\rm sat}$, where $\rho_{\rm sat}=2.7\times10^{14}\text{g/cm}^3$ is the nuclear saturation density. As described in Sec.~\ref{sec:methods}, the EoS is fixed at densities below $0.5 \rho_{\rm sat}$ to SFHo. We list this anchoring pressure as well in Table~\ref{table:EOS_params} for convenience. For details on constructing piecewise polytropic EoSs using these parameters, see, e.g. \citet{Read:2008iy, Ozel:2009da, Raithel:2016bux}.

\begin{deluxetable}{llllllllll}
\tabletypesize{\footnotesize}
\tablewidth{0.6\textwidth} 
\tablecaption{\label{table:EOS_params} Parameters describing the three new piecewise, polytropic EoSs considered in this work. The first column gives the model labels from Fig.~\ref{fig:newEOS}. The second and third columns report the radius and tidal deformability of a 1.4~$\Ms$ neutron star, respectively. The remaining columns report the pressures at each fiducial density in the piecewise polytropic model. All pressures are given in units of dyn/cm$^2$.} 
\tablehead{
\colhead{ Model }  &
\colhead{$R_{1.4}$ (km)  }  &
\colhead{$\Lambda_{1.4}$ }  &
\colhead{$P(0.5 \rho_{\rm sat})$}  &
\colhead{$P(0.86\rho_{\rm sat})$}  &
\colhead{$P(1.47\rho_{\rm sat})$} &
\colhead{$P(2.52\rho_{\rm sat})$ }  &
\colhead{$P(4.32\rho_{\rm sat})$}  &
\colhead{$P(7.4\rho_{\rm sat})$}  
}
\startdata   
A &  10.8  & 216 &  8.88$\times 10^{32}$ & 1.80$\times 10^{33}$  & 4.00$\times 10^{33}$ & 4.10$\times 10^{34}$ & 5.20$\times 10^{35}$ & 1.60$\times 10^{36}$ \\
B  & 13.0 & 608 &  8.88$\times 10^{32}$ & 4.00$\times10^{33}$ & 1.55$\times10^{34}$ & 1.10$ \times 10^{35}$ & 4.00$\times10^{35}$ & 1.00$\times10^{36}$ \\
C &  13.0 & 665 &   8.88$\times 10^{32}$ & 3.10$\times 10^{33}$ & 1.43$\times 10^{34}$ & 1.80$\times 10^{35}$ & 3.00$\times 10^{35}$ & 6.00$\times 10^{35}$\\
 \enddata
\end{deluxetable}

\section{Alternate definitions of the mass-radius slope}

\centerwidetable
\begin{deluxetable*}{llllllllll}
\tabletypesize{\footnotesize}
\tablewidth{0.6\textwidth} 
\tablecaption{\label{table:slopes} Fit coefficients for eq.~(\ref{eq:two_param}) for alternate definitions of the mass-radius slope. The formatting is otherwise identical to Table~\ref{table:fits}. The results with the slope parameter $R_{1.4}/R_{1.8}$ are repeated for reference.} 
\tablehead{
\colhead{ Slope }  &
\colhead{ $R_x$ }  &
\colhead{$b_0$  }  &
\colhead{$b_1$ }  &
\colhead{$b_2$}  &
\colhead{$b_3$}  &
\colhead{Adjusted $\mathcal{R}^2$} &
\colhead{BIC }  &
\colhead{Max resid (kHz) }  &
\colhead{Mean resid (kHz) }  
}
\startdata   
     &	 $R_{1.4}$   & 0.081  & -0.592 & 0.007 & 8.900 & 0.927 & -20.2 & 0.21& 0.08\T\\
    & $R_{1.6}$   & 5.403 & -1.083 & 0.027 & 6.544 & 0.923 & -19.3 & 0.21& 0.08\\
$R_{1.4}/R_{1.8}$    & $R_{1.8}$   & 12.179 & -1.821 & 0.057 & 4.287 & 0.933 & -21.7 & 0.20& 0.07\\
 &  $R_{2.0}$   & 19.455 & -2.459 & 0.083 & 0.837 & 0.929 & -20.6 & 0.30& 0.07\\
 &  $R_{\rm max}$   & 21.421 & -2.871 & 0.104 & 0.504 & 0.913 & -17.2 & 0.23& 0.08\B\\ 
 \hline
  & $R_{1.4}$   & 16.073 & -1.780 & 0.057 & -0.204 & 0.848 & -7.7 & 0.44& 0.10\T\\
 & $R_{1.6}$   & 16.180  & -1.792  & 0.057  & -0.148  & 0.874  & -10.9  & 0.39 & 0.09\\
$\frac{\partial R}{\partial M} \biggr\rvert_{1.2}$   & $R_{1.8}$   & 18.674  & -2.191  & 0.073  & -0.102  & 0.918  & -18.1  & 0.28 & 0.08\\
& $R_{2.0}$   & 22.530  & -2.848  & 0.100  & -0.066  & 0.946  & -25.2  & 0.27 & 0.06\\
 & $R_{\rm max}$   & 23.182  & -3.112  & 0.116  & -0.070  & 0.937  & -22.7  & 0.21 & 0.07\B\\
  \hline
&  $R_{1.4}$   & 10.058  & -0.790  & 0.016  & -0.244  & 0.880  & -11.7  & 0.34 & 0.09\T\\
& $R_{1.6}$   & 12.202  & -1.139  & 0.030  & -0.177  & 0.889  & -13.0  & 0.33 & 0.09\\
$\frac{\partial R}{\partial M} \biggr\rvert_{1.4}$   & $R_{1.8}$   & 16.467  & -1.828  & 0.058  & -0.118  & 0.919  & -18.4  & 0.25 & 0.08\\
& $R_{2.0}$   & 21.120  & -2.606  & 0.090  & -0.049  & 0.934  & -21.8  & 0.29 & 0.06\\
& $R_{\rm max}$   & 22.746  & -3.030  & 0.112  & -0.057  & 0.923  & -19.3  & 0.24 & 0.08\B\\   
\hline 
	& $R_{1.4}$   & 7.552  & -0.365  & -0.002  & -0.283  & 0.910  & -16.6  & 0.23 & 0.08\T\\
  & $R_{1.6}$   & 11.070  & -0.944  & 0.022  & -0.206  & 0.910  & -16.6  & 0.25 & 0.08\\
$\frac{\partial R}{\partial M} \biggr\rvert_{1.6}$ &  $R_{1.8}$   & 16.052  & -1.755  & 0.055  & -0.133  & 0.925  & -19.7  & 0.20 & 0.07\\
 & $R_{2.0}$   & 20.285  & -2.457  & 0.083  & -0.025  & 0.928  & -20.4  & 0.30 & 0.07\\
& $R_{\rm max}$   & 21.976  & -2.880  & 0.105  & -0.017  & 0.913  & -17.2  & 0.23 & 0.08\B\\ 
\enddata
\end{deluxetable*}

In this Letter, we have shown that the existing family of single-parameter, 
quasi-universal relations between $f_2$ and the neutron star radius 
break down for EoSs that predict non-vertical slopes in their mass-radius relations,
but that universality may be restored by adding in a second parameter that depends on
the slope. In this appendix, we confirm that this conclusion is robust to different definitions
of the slope, including the value of $\partial R/\partial M$ evaluated at
either 1.2, 1.4, or 1.6~$\Ms$. We report the statistics for these fits in Table~\ref{table:slopes}.

In particular, for each of these alternate definitions of the slope, there 
is still strong evidence in favor of adding a second parameter to the relationship between
$f_2$ and $R_{1.4}$,$R_{1.6}$, or $R_{1.8}$ ($\Delta$ BIC $\gtrsim$ 5, compared to the
mono-parametric fits reported in Table~\ref{table:fits}). As was found in the main text, for each these
definitions the evidence favoring a second parameter again decreases 
as $f_2$ is correlated with radii at increasingly large masses, such that
for $f_2(R_{2.0})$, there is no significant evidence in favor of adding a second parameter.
This confirms that the conclusions of the main Letter do not change depending
on where the slope is defined.

However, in spite of these qualitatively consistent trends, the goodness of fit
does change slightly depending on the definition of the slope parameter.
For example, in the fit for $f_2(R_{1.4}, R_{1.4}/R_{1.8})$ reported in the main
text, the adjusted $\mathcal{R}^2$ was 0.927. For $f_2(R_{1.4}, \partial R/\partial M)$,
the adjusted $\mathcal{R}^2$ value is reduced, ranging from 0.85 to 0.91, depending
on the mass at which the slope is evaluated.

\begin{figure*}[!ht]
\centering
\includegraphics[width=0.86\textwidth]{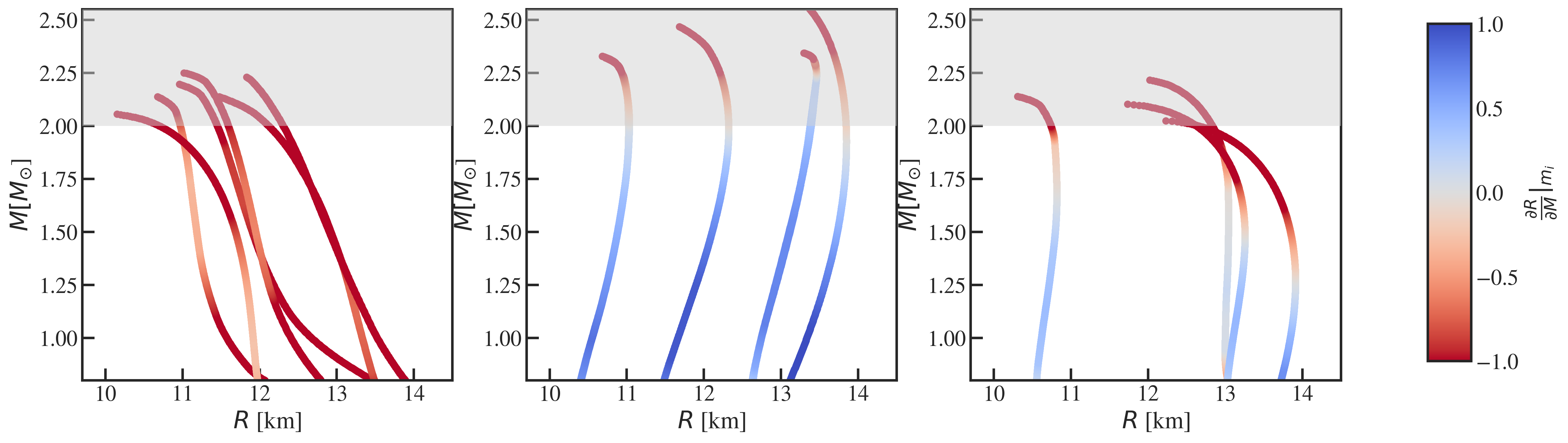}
\caption{\label{fig:MRslopes} Mass-radius relation for the same set of EoSs shown in Fig.~\ref{fig:EoS},
but now color-coded according to the instantaneous slope. The models have been divided into three panels
for visual clarity. Masses above 2~$\Ms$ have been grayed out because the mass-radius curves of all EoSs
turn-over near the maximum mass, and so the changing slope in this mass range is not especially interesting. 
Focusing instead then on the mass range of 1-2~$\Ms$, it is clear that
the majority of these curves can be well defined by a single slope parameter;
however, for a subset of cases (grouped in the right-most panel), the slope is sensitive to the mass at which it is defined.}
\end{figure*}

We find that the sensitivity to the definition of the mass-radius slope comes primarily
from a subset of models, as can be seen in Fig.~\ref{fig:MRslopes}.
Figure~\ref{fig:MRslopes} shows the same set of EoSs from Fig.~\ref{fig:EoS},
but now color-coded by a local (or instantaneous) slope. The models have been
divided into three panels for visual clarity, with forwards-tilting MR curves
in the left panel, backwards-bending models in the middle panel, and models
with varying mass-radius slope in the right panel. While the majority
of curves can be well represented by a single slope parameter acorss the mass range
of interest, a subset of models (grouped in the right panel of Fig.~\ref{fig:MRslopes})
show some degree of curvature in their mass-radius relations.
These include the finite-temperature EoSs BHB$\Lambda\phi$ and TMA,
as well as two of the parametric models. As a result, the goodness-of-fit for the
quasi-universal relation relating $f_2$ to a characteristic radius and the mass-radius
slope can be sensitive to where, precisely, the slope is defined.

In summary, we find that the main conclusions of this Letter are robust to these alternate definitions of the
slope. However, for broader families of EoSs, it may be necessary
to revisit this definition, or even to include additional correction
terms to the quasi-universal relations that describe the shape of the mass-radius
relation at beyond linear order. Understanding this dependence will require
additional simulations, spanning a wide range of both linear
and curved mass-radius relations, which expands the dimensionality of 
the problem significantly. We leave such an investigation to a future study.

\bibliographystyle{yahapj}
\bibliography{main,non_inspire}

\begin{thebibliography}{}
\providecommand\natexlab[1]{#1}
\providecommand\JournalTitle[1]{#1}

\bibitem[{Abbott {et~al.}(2017)}]{TheLIGOScientific:2017qsa}
Abbott, B.~P., {et~al.} 2017,
  \href{http://dx.doi.org/10.1103/PhysRevLett.119.161101}{\JournalTitle{Phys.
  Rev. Lett.}, 119, 161101}

\bibitem[{Antoniadis {et~al.}(2013)}]{Antoniadis:2013pzd}
Antoniadis, J., {et~al.} 2013,
  \href{http://dx.doi.org/10.1126/science.1233232}{\JournalTitle{Science}, 340,
  6131}

\bibitem[{Baiotti(2019)}]{Baiotti:2019sew}
Baiotti, L. 2019,
  \href{http://dx.doi.org/10.1016/j.ppnp.2019.103714}{\JournalTitle{Prog. Part.
  Nucl. Phys.}, 109, 103714}

\bibitem[{Baiotti \& Rezzolla(2017)}]{Baiotti:2016qnr}
Baiotti, L., \& Rezzolla, L. 2017,
  \href{http://dx.doi.org/10.1088/1361-6633/aa67bb}{\JournalTitle{Rept. Prog.
  Phys.}, 80, 096901}

\bibitem[{Banik {et~al.}(2014)Banik, Hempel, \& Bandyopadhyay}]{Banik:2014qja}
Banik, S., Hempel, M., \& Bandyopadhyay, D. 2014,
  \href{http://dx.doi.org/10.1088/0067-0049/214/2/22}{\JournalTitle{Astrophys.
  J. Suppl.}, 214, 22}

\bibitem[{Bauswein {et~al.}(2019)Bauswein, Bastian, Blaschke, Chatziioannou,
  Clark, Fischer, \& Oertel}]{Bauswein:2018bma}
Bauswein, A., Bastian, N.-U.~F., Blaschke, D.~B., {et~al.} 2019,
  \href{http://dx.doi.org/10.1103/PhysRevLett.122.061102}{\JournalTitle{Phys.
  Rev. Lett.}, 122, 061102}

\bibitem[{Bauswein \& Janka(2012)}]{Bauswein:2011tp}
Bauswein, A., \& Janka, H.~T. 2012,
  \href{http://dx.doi.org/10.1103/PhysRevLett.108.011101}{\JournalTitle{Phys.
  Rev. Lett.}, 108, 011101}

\bibitem[{Bauswein {et~al.}(2012)Bauswein, Janka, Hebeler, \&
  Schwenk}]{Bauswein:2012ya}
Bauswein, A., Janka, H.~T., Hebeler, K., \& Schwenk, A. 2012,
  \href{http://dx.doi.org/10.1103/PhysRevD.86.063001}{\JournalTitle{Phys. Rev.
  D}, 86, 063001}

\bibitem[{Bauswein \& Stergioulas(2019)}]{Bauswein:2019ybt}
Bauswein, A., \& Stergioulas, N. 2019,
  \href{http://dx.doi.org/10.1088/1361-6471/ab2b90}{\JournalTitle{J. Phys. G},
  46, 113002}

\bibitem[{Bernuzzi(2020)}]{Bernuzzi:2020tgt}
Bernuzzi, S. 2020,
  \href{http://dx.doi.org/10.1007/s10714-020-02752-5}{\JournalTitle{Gen. Rel.
  Grav.}, 52, 108}

\bibitem[{Bernuzzi {et~al.}(2015)Bernuzzi, Dietrich, \&
  Nagar}]{Bernuzzi:2015rla}
Bernuzzi, S., Dietrich, T., \& Nagar, A. 2015,
  \href{http://dx.doi.org/10.1103/PhysRevLett.115.091101}{\JournalTitle{Phys.
  Rev. Lett.}, 115, 091101}

\bibitem[{Chatziioannou(2020)}]{Chatziioannou:2020pqz}
Chatziioannou, K. 2020,
  \href{http://dx.doi.org/10.1007/s10714-020-02754-3}{\JournalTitle{Gen. Rel.
  Grav.}, 52, 109}

\bibitem[{Cromartie {et~al.}(2019)}]{NANOGrav:2019jur}
Cromartie, H.~T., {et~al.} 2019,
  \href{http://dx.doi.org/10.1038/s41550-019-0880-2}{\JournalTitle{Nature
  Astron.}, 4, 72}

\bibitem[{Demorest {et~al.}(2010)Demorest, Pennucci, Ransom, Roberts, \&
  Hessels}]{Demorest:2010bx}
Demorest, P., Pennucci, T., Ransom, S., Roberts, M., \& Hessels, J. 2010,
  \href{http://dx.doi.org/10.1038/nature09466}{\JournalTitle{Nature}, 467,
  1081}

\bibitem[{Duez {et~al.}(2005)Duez, Liu, Shapiro, \& Stephens}]{Duez:2005sf}
Duez, M.~D., Liu, Y.~T., Shapiro, S.~L., \& Stephens, B.~C. 2005,
  \href{http://dx.doi.org/10.1103/PhysRevD.72.024028}{\JournalTitle{Phys. Rev.
  D}, 72, 024028}

\bibitem[{Etienne {et~al.}(2015)Etienne, Paschalidis, Haas, M\"osta, \&
  Shapiro}]{Etienne:2015cea}
Etienne, Z.~B., Paschalidis, V., Haas, R., M\"osta, P., \& Shapiro, S.~L. 2015,
  \href{http://dx.doi.org/10.1088/0264-9381/32/17/175009}{\JournalTitle{Class.
  Quant. Grav.}, 32, 175009}

\bibitem[{Fonseca {et~al.}(2016)}]{Fonseca:2016tux}
Fonseca, E., {et~al.} 2016,
  \href{http://dx.doi.org/10.3847/0004-637X/832/2/167}{\JournalTitle{Astrophys.
  J.}, 832, 167}

\bibitem[{Gandolfi {et~al.}(2014)Gandolfi, Carlson, Reddy, Steiner, \&
  Wiringa}]{Gandolfi:2013baa}
Gandolfi, S., Carlson, J., Reddy, S., Steiner, A.~W., \& Wiringa, R.~B. 2014,
  \href{http://dx.doi.org/10.1140/epja/i2014-14010-5}{\JournalTitle{Eur. Phys.
  J. A}, 50, 10}

\bibitem[{Gourgoulhon {et~al.}(2001)Gourgoulhon, Grandclement, Taniguchi,
  Marck, \& Bonazzola}]{Gourgoulhon:2000nn}
Gourgoulhon, E., Grandclement, P., Taniguchi, K., Marck, J.-A., \& Bonazzola,
  S. 2001,
  \href{http://dx.doi.org/10.1103/PhysRevD.63.064029}{\JournalTitle{Phys. Rev.
  D}, 63, 064029}

\bibitem[{Guerra~Chaves \& Hinderer(2019)}]{GuerraChaves:2019foa}
Guerra~Chaves, A., \& Hinderer, T. 2019,
  \href{http://dx.doi.org/10.1088/1361-6471/ab45be}{\JournalTitle{J. Phys. G},
  46, 123002}

\bibitem[{Hempel \& Schaffner-Bielich(2010)}]{Hempel:2009mc}
Hempel, M., \& Schaffner-Bielich, J. 2010,
  \href{http://dx.doi.org/10.1016/j.nuclphysa.2010.02.010}{\JournalTitle{Nucl.
  Phys. A}, 837, 210}

\bibitem[{Hilditch {et~al.}(2013)Hilditch, Bernuzzi, Thierfelder, Cao, Tichy,
  \& Bruegmann}]{Hilditch:2012fp}
Hilditch, D., Bernuzzi, S., Thierfelder, M., {et~al.} 2013,
  \href{http://dx.doi.org/10.1103/PhysRevD.88.084057}{\JournalTitle{Phys. Rev.
  D}, 88, 084057}

\bibitem[{Hunter(2007)}]{Hunter:2007}
Hunter, J.~D. 2007,
  \href{http://dx.doi.org/10.1109/MCSE.2007.55}{\JournalTitle{Computing in
  Science \& Engineering}, 9, 90}

\bibitem[{Jeffreys(1961)}]{Jeffreys61}
Jeffreys, H. 1961, Theory of Probability, 3rd edn. (Oxford, England: Oxford)

\bibitem[{Lattimer \& Swesty(1991)}]{Lattimer:1991nc}
Lattimer, J.~M., \& Swesty, F.~D. 1991,
  \href{http://dx.doi.org/10.1016/0375-9474(91)90452-C}{\JournalTitle{Nucl.
  Phys. A}, 535, 331}

\bibitem[{Liddle(2007)}]{Liddle:2007fy}
Liddle, A.~R. 2007,
  \href{http://dx.doi.org/10.1111/j.1745-3933.2007.00306.x}{\JournalTitle{Mon.
  Not. Roy. Astron. Soc.}, 377, L74}

\bibitem[{Loffler {et~al.}(2012)}]{Loffler:2011ay}
Loffler, F., {et~al.} 2012,
  \href{http://dx.doi.org/10.1088/0264-9381/29/11/115001}{\JournalTitle{Class.
  Quant. Grav.}, 29, 115001}

\bibitem[{McLerran \& Reddy(2019)}]{McLerran:2018hbz}
McLerran, L., \& Reddy, S. 2019,
  \href{http://dx.doi.org/10.1103/PhysRevLett.122.122701}{\JournalTitle{Phys.
  Rev. Lett.}, 122, 122701}

\bibitem[{Miller {et~al.}(2021)}]{Miller:2021qha}
Miller, M.~C., {et~al.} 2021,
  \href{http://dx.doi.org/10.3847/2041-8213/ac089b}{\JournalTitle{Astrophys. J.
  Lett.}, 918, L28}

\bibitem[{Most {et~al.}(2019)Most, Papenfort, \& Rezzolla}]{Most:2019kfe}
Most, E.~R., Papenfort, L.~J., \& Rezzolla, L. 2019,
  \href{http://dx.doi.org/10.1093/mnras/stz2809}{\JournalTitle{Mon. Not. Roy.
  Astron. Soc.}, 490, 3588}

\bibitem[{Most \& Raithel(2021)}]{Most:2021ktk}
Most, E.~R., \& Raithel, C.~A. 2021,
  \href{http://dx.doi.org/10.1103/PhysRevD.104.124012}{\JournalTitle{Phys. Rev.
  D}, 104, 124012}

\bibitem[{\"Ozel \& Freire(2016)}]{Ozel:2016oaf}
\"Ozel, F., \& Freire, P. 2016,
  \href{http://dx.doi.org/10.1146/annurev-astro-081915-023322}{\JournalTitle{Ann.
  Rev. Astron. Astrophys.}, 54, 401}

\bibitem[{Ozel \& Psaltis(2009)}]{Ozel:2009da}
Ozel, F., \& Psaltis, D. 2009,
  \href{http://dx.doi.org/10.1103/PhysRevD.80.103003}{\JournalTitle{Phys. Rev.
  D}, 80, 103003}

\bibitem[{Ozel {et~al.}(2016)Ozel, Psaltis, Guver, Baym, Heinke, \&
  Guillot}]{Ozel:2015fia}
Ozel, F., Psaltis, D., Guver, T., {et~al.} 2016,
  \href{http://dx.doi.org/10.3847/0004-637X/820/1/28}{\JournalTitle{Astrophys.
  J.}, 820, 28}

\bibitem[{Paschalidis \& Stergioulas(2017)}]{Paschalidis:2016vmz}
Paschalidis, V., \& Stergioulas, N. 2017,
  \href{http://dx.doi.org/10.1007/s41114-017-0008-x}{\JournalTitle{Living Rev.
  Rel.}, 20, 7}

\bibitem[{Raaijmakers {et~al.}(2021)Raaijmakers, Greif, Hebeler, Hinderer,
  Nissanke, Schwenk, Riley, Watts, Lattimer, \& Ho}]{Raaijmakers:2021uju}
Raaijmakers, G., Greif, S.~K., Hebeler, K., {et~al.} 2021,
  \href{http://dx.doi.org/10.3847/2041-8213/ac089a}{\JournalTitle{Astrophys. J.
  Lett.}, 918, L29}

\bibitem[{Radice {et~al.}(2020)Radice, Bernuzzi, \& Perego}]{Radice:2020ddv}
Radice, D., Bernuzzi, S., \& Perego, A. 2020,
  \href{http://dx.doi.org/10.1146/annurev-nucl-013120-114541}{\JournalTitle{Ann.
  Rev. Nucl. Part. Sci.}, 70, 95}

\bibitem[{Raithel {et~al.}(2022)Raithel, Espino, \&
  Paschalidis}]{Raithel:2022nab}
Raithel, C., Espino, P., \& Paschalidis, V. 2022,
  \href{http://arxiv.org/abs/2206.14838}{{\sffamily arXiv:2206.14838
  [astro-ph.HE]}}

\bibitem[{Raithel(2019)}]{Raithel:2019uzi}
Raithel, C.~A. 2019,
  \href{http://dx.doi.org/10.1140/epja/i2019-12759-5}{\JournalTitle{Eur. Phys.
  J. A}, 55, 80}

\bibitem[{Raithel {et~al.}(2016)Raithel, Ozel, \& Psaltis}]{Raithel:2016bux}
Raithel, C.~A., Ozel, F., \& Psaltis, D. 2016,
  \href{http://dx.doi.org/10.3847/0004-637X/831/1/44}{\JournalTitle{Astrophys.
  J.}, 831, 44}

\bibitem[{Raithel {et~al.}(2017)Raithel, \"Ozel, \& Psaltis}]{Raithel:2017ity}
Raithel, C.~A., \"Ozel, F., \& Psaltis, D. 2017,
  \href{http://dx.doi.org/10.3847/1538-4357/aa7a5a}{\JournalTitle{Astrophys.
  J.}, 844, 156}

\bibitem[{Raithel {et~al.}(2019)Raithel, Ozel, \& Psaltis}]{Raithel:2019gws}
Raithel, C.~A., Ozel, F., \& Psaltis, D. 2019,
  \href{http://dx.doi.org/10.3847/1538-4357/ab08ea}{\JournalTitle{Astrophys.
  J.}, 875, 12}

\bibitem[{Read {et~al.}(2009)Read, Lackey, Owen, \& Friedman}]{Read:2008iy}
Read, J.~S., Lackey, B.~D., Owen, B.~J., \& Friedman, J.~L. 2009,
  \href{http://dx.doi.org/10.1103/PhysRevD.79.124032}{\JournalTitle{Phys. Rev.
  D}, 79, 124032}

\bibitem[{Rezzolla \& Takami(2016)}]{Rezzolla:2016nxn}
Rezzolla, L., \& Takami, K. 2016,
  \href{http://dx.doi.org/10.1103/PhysRevD.93.124051}{\JournalTitle{Phys. Rev.
  D}, 93, 124051}

\bibitem[{Riley {et~al.}(2021)}]{Riley:2021pdl}
Riley, T.~E., {et~al.} 2021,
  \href{http://dx.doi.org/10.3847/2041-8213/ac0a81}{\JournalTitle{Astrophys. J.
  Lett.}, 918, L27}

\bibitem[{Schneider {et~al.}(2017)Schneider, Roberts, \&
  Ott}]{Schneider:2017tfi}
Schneider, A.~S., Roberts, L.~F., \& Ott, C.~D. 2017,
  \href{http://dx.doi.org/10.1103/PhysRevC.96.065802}{\JournalTitle{Phys. Rev.
  C}, 96, 065802}

\bibitem[{Schnetter {et~al.}(2004)Schnetter, Hawley, \&
  Hawke}]{Schnetter:2003rb}
Schnetter, E., Hawley, S.~H., \& Hawke, I. 2004,
  \href{http://dx.doi.org/10.1088/0264-9381/21/6/014}{\JournalTitle{Class.
  Quant. Grav.}, 21, 1465}

\bibitem[{Steiner {et~al.}(2013)Steiner, Hempel, \& Fischer}]{Steiner:2012rk}
Steiner, A.~W., Hempel, M., \& Fischer, T. 2013,
  \href{http://dx.doi.org/10.1088/0004-637X/774/1/17}{\JournalTitle{Astrophys.
  J.}, 774, 17}

\bibitem[{Steiner {et~al.}(2010)Steiner, Lattimer, \& Brown}]{Steiner:2010fz}
Steiner, A.~W., Lattimer, J.~M., \& Brown, E.~F. 2010,
  \href{http://dx.doi.org/10.1088/0004-637X/722/1/33}{\JournalTitle{Astrophys.
  J.}, 722, 33}

\bibitem[{Steiner {et~al.}(2016)Steiner, Lattimer, \& Brown}]{Steiner:2015aea}
---. 2016,
  \href{http://dx.doi.org/10.1140/epja/i2016-16018-1}{\JournalTitle{Eur. Phys.
  J. A}, 52, 18}

\bibitem[{Stergioulas {et~al.}(2011)Stergioulas, Bauswein, Zagkouris, \&
  Janka}]{Stergioulas:2011gd}
Stergioulas, N., Bauswein, A., Zagkouris, K., \& Janka, H.-T. 2011,
  \href{http://dx.doi.org/10.1111/j.1365-2966.2011.19493.x}{\JournalTitle{Mon.
  Not. Roy. Astron. Soc.}, 418, 427}

\bibitem[{Takami {et~al.}(2014)Takami, Rezzolla, \& Baiotti}]{Takami:2014zpa}
Takami, K., Rezzolla, L., \& Baiotti, L. 2014,
  \href{http://dx.doi.org/10.1103/PhysRevLett.113.091104}{\JournalTitle{Phys.
  Rev. Lett.}, 113, 091104}

\bibitem[{Takami {et~al.}(2015)Takami, Rezzolla, \& Baiotti}]{Takami:2014tva}
---. 2015,
  \href{http://dx.doi.org/10.1103/PhysRevD.91.064001}{\JournalTitle{Phys. Rev.
  D}, 91, 064001}

\bibitem[{Tan {et~al.}(2021)Tan, Dexheimer, Noronha-Hostler, \&
  Yunes}]{Tan:2021nat}
Tan, H., Dexheimer, V., Noronha-Hostler, J., \& Yunes, N. 2021,
  \href{http://arxiv.org/abs/2111.10260}{{\sffamily arXiv:2111.10260
  [astro-ph.HE]}}

\bibitem[{Toki {et~al.}(1995)Toki, Hirata, Sugahara, Sumiyoshi, \&
  Tanihata}]{Toki:1995ya}
Toki, H., Hirata, D., Sugahara, Y., Sumiyoshi, K., \& Tanihata, I. 1995,
  \href{http://dx.doi.org/10.1016/0375-9474(95)00161-S}{\JournalTitle{Nucl.
  Phys. A}, 588, c357}

\bibitem[{Torres-Rivas {et~al.}(2019)Torres-Rivas, Chatziioannou, Bauswein, \&
  Clark}]{Torres-Rivas:2018svp}
Torres-Rivas, A., Chatziioannou, K., Bauswein, A., \& Clark, J.~A. 2019,
  \href{http://dx.doi.org/10.1103/PhysRevD.99.044014}{\JournalTitle{Phys. Rev.
  D}, 99, 044014}

\bibitem[{Vretinaris {et~al.}(2020)Vretinaris, Stergioulas, \&
  Bauswein}]{Vretinaris:2019spn}
Vretinaris, S., Stergioulas, N., \& Bauswein, A. 2020,
  \href{http://dx.doi.org/10.1103/PhysRevD.101.084039}{\JournalTitle{Phys. Rev.
  D}, 101, 084039}

\bibitem[{Waskom(2021)}]{Waskom2021}
Waskom, M.~L. 2021,
  \href{http://dx.doi.org/10.21105/joss.03021}{\JournalTitle{Journal of Open
  Source Software}, 6, 3021}

\bibitem[{Yagi \& Yunes(2016)}]{Yagi:2015pkc}
Yagi, K., \& Yunes, N. 2016,
  \href{http://dx.doi.org/10.1088/0264-9381/33/13/13LT01}{\JournalTitle{Class.
  Quant. Grav.}, 33, 13LT01}

\end{thebibliography}

\end{document}